# Self-assembly of parallel atomic wires and periodic clusters of silicon on a vicinal Si(111) surface

Takeharu Sekiguchi, Shunji Yoshida, and Kohei M. Itoh

*Department of Applied Physics and Physico-Informatics and CREST-JST, Keio University, Yokohama 223-8522, Japan*

Silicon self-assembly at step edges in the initial stage of homoepitaxial growth on a vicinal Si(111) surface is studied by scanning tunneling microscopy (STM). The resulting atomic structures change dramatically from a parallel array of 0.7 nm wide wires to one dimensionally aligned periodic clusters of the diameter ~ 2 nm and periodicity 2.7 nm in the very narrow range of growth temperatures between 400 and 300 °C. These nanostructures are expected to play an important role in future development of silicon quantum computers. Mechanisms leading to such distinct structures are discussed.

68.65.La, 68.35.Bs, 81.15.Hi, 03.67.Lx

Atomic-level understandings of the silicon surface reconstructions and crystal growth processes are of great interest to scientists and engineers. In particular, Si(111) has been studied extensively since the structural asymmetry between the two half-unit cells of the 7×7 reconstruction of the dimer–adatom–stacking-fault (DAS) structure [1] leads to a wealth of interesting behaviors especially when different atoms are deposited onto this surface [2]. Moreover, thanks to the stiffness of the steps on the 7×7-reconstructed surface, a parallel array of very straight step edges with the density of kinks (having the width of the 7×7 unit cell) as small as 0.12 $\mu m^{-1}$ has been successfully achieved on its vicinal surface [3,4]. Such straight step-edge surfaces have been proven useful as templates for fabrication of a variety of nanostructures for solid-state research [5], e.g., one-dimensional gold structures formed on such templates are found to preserve metallicity due to the fractional band filling of strongly correlated electrons [6].

In this Letter, we report self-assembly of deposited Si atoms at the atomically straight steps on a vicinal Si(111) surface, i.e., silicon nanostructure formation at the very initial stage of silicon homoepitaxy. We have been able to prepare an array of atomically straight step edges of micrometer lengths. They are free of not only the single 7×7 wide kinks but also of single atomic-level kinks [7]. As a result, every straight step edge in this array has an identical atomic structure, and, therefore, Si atoms deposited onto it can be distinctly identified from the substrate Si atoms by scanning tunneling microscopy (STM). We find that at the very beginning of the growth, either a parallel array of atomic wires or a one-dimensional periodic array of 2-nm wide clusters is formed by the deposited Si atoms in the narrow temperature range 400–300 °C. Both the atomic wires and the periodic clusters are expected to play important roles in the future developments of silicon based nuclear spin quantum computers [8–10]. We have shown that each nuclear spin of $^{29}$Si isotopes arranged in an atomic wire on a nuclear-spin-free $^{28}$Si substrate is an excellent candidate as a qubit [11]. Also, a one-dimensional periodic chain of $^{29}$Si nuclear-spin clusters on a nuclear-spin-free $^{28}$Si substrate can serve as a building block in the recently proposed spin cluster quantum computing [12,13]. The present investigation, therefore, is important not only for better understanding of the very initial stage of silicon homoepitaxial growth but also towards realization of quantum computers.

Our method to prepare the atomic-kink-free straight step template is described in Ref. [7]. In short, a commercially available p-type Si(111) wafer is polished to make a polar miscut of 1° towards the [$\bar{1}\bar{1}2$] direction with an intentional azimuthal misorientation of about 3°. Thermal treatment of this substrate is performed in an ultrahigh vacuum chamber equipped with an STM (Omicron UHV STM-1) and a molecular beam epitaxy (MBE) system. After 1260 °C flash cleaning, the substrate is quenched to 800 °C or room temperature in a few seconds to skip the coexisting phase between the "1×1" and the 7×7 reconstructions around 860 °C. Then, the substrate is annealed at 800 °C for 10 hours with DC current in the step-parallel direction that ascends the kinks ("kink-up" direction), to induce a specific electromigration of surface atoms that leads to the step straightening. After slow cool-down to room temperature, the straight step structure is confirmed by STM. Further details including a proposed mechanism of step straightening are discussed in Ref. [7]. In the following homoepitaxial growth, Si is evaporated from a solid Si source in an effusion cell. The source temperature is held at 1150 °C during the deposition for a slow growth rate of $5.6\times10^{-5}$ BL/s (BL or bilayer is 0.314 nm in film thickness). For the convenience, we define "nominal step coverage" $w$ as the average width of 1-BL high island formed at each step edge in step-flow growth mode: $w = \Theta W$, where $\Theta$ and $W$ are the film thickness in units of BL and the average terrace width, respectively. $W$ is 18 nm for the 1° polar miscut. The substrate temperature is monitored by the DC current calibrated against the temperature measured by an infrared pyrometer. The pyrometer itself has been calibrated by RHEED pattern change at the phase transition temperature 860 °C. All the STM images presented in this letter are taken at room temperature.

Figure 1(a) shows the atomically straight step-edge templates of the one BL step height before homoepitaxial growth. The kink density, the number of kinks in a given step length, we achieve is less than one per 2.5 μm or $1.6\times10^{-4}/a$, where $a = 0.38$ nm is the side length of the 1×1 half-unit cell triangle. More importantly, every straight step edge after our substrate preparation has an identical atomic structure, so that Si atoms that will be deposited later can be identified from the substrate atoms. A high-resolution STM image of such a step and a corresponding structure model are shown in Figs. 2(a) and 2(b), respectively. While the upper terrace is terminated with the dimer wall at the 7×7 unit cell boundary (indicated by green arrows) as reported for most kinds of $[\bar{1}\,\bar{1}\,2]$ (hereafter $U$-type) steps [14], a "transition region" containing four adatoms per $7a$ length is formed between the step edge (the green arrows) and the dimer wall (blue arrows) on the lower terrace. The displacement of the 7×7 structure across the step edge in the step-perpendicular direction, i.e., the width of the transition region, is *always* $(2 + 2/3)b = 0.89$ nm, while there is no shift in the step-parallel direction. Here $b = 0.33$ nm is the lateral height of the 1×1 half-unit cell triangle. This step configuration is classified as U(2,0) [15], where the notation $U(n,m)$ represents the shift of the $(n + 2/3)b$ perpendicular and $ma$ parallel to the $U$-type step [16].

A sub-BL amount of Si atoms is deposited on such a template surface at a specific growth temperature $T_g$. At $T_g = 700$ °C, the step edges extend to their perpendicular direction in units of the 7×7 unit cell, i.e., 7*b* width in one BL thickness, as reported previously [17,18]. This leads to successive rebuilding of the $U$(2,0) structures.

The growth mode changes dramatically when $T_g$ is lowered to 400 °C. Initially, each step edge extends only by the width 2*b*, i.e., a single row of Si adatoms is formed along the step edge as shown in the STM image of Fig. 1(b). Even when *w* is increased up to 7*b*, these atomic rows survives while defective 7×7 structure begins to form next to some of them [19]. An empty-state STM image and a corresponding structure model of the single adatom row are shown in Figs. 2(c) and 2(d), respectively. Comparison of this image with Fig. 2(a) shows clearly that the Si adatoms are arranged with regular spacing 2*a* in a straight line along the lower step edge. From the rest atom positions in the corresponding filled-state STM image [Fig. 2(e)], it is clear that the single-wire region is unfaulted and the new step edge is terminated by the dimer wall. The unfaulted stacking in the transition region of the lower terrace allows the formation of the adatom rows while the faulted stacking beyond the transition region suppresses further row-by-row growth, as suggested previously [16]. Moreover, the definite width of the transition region is crucial for exclusive formation of the single adatom row since the width (2 + 2/3)*b* is large enough to form the single row (2*b*) but too small to grow a double row (4*b*) or other wider structures. Otherwise, if the transition width were larger than 3*b* = 1.00 nm, the faulted halves of the 3×3 structure (3*b* wide) would be formed instead of the single adatom row [20]. In this regard, the $U$(2,0) structure at every step edge we have successfully prepared is serendipitously suitable for the exclusive formation of the single Si adatom rows.

Another form of unexpected atomic structures of deposited Si is obtained by reduction of the growth temperature only by 100 °C. Figure 1(c) is an STM image of a sample of $w = 7b$ and $T_g = 300$ °C that shows a one-dimensional (1D) array of clusters self-assembled along the upper step edges. Each cluster is about 2 nm in diameter and less than a half BL in height. The magnified image in Fig. 3 shows clearly that the clusters occupy preferentially the unfaulted halves at the upper step edges, resulting in a 1D arrangement with the periodicity of $7a$ (2.7 nm). This is interesting because a wide variety of atomic species deposited on a flat Si(111) terrace nucleates preferentially on the *faulted* halves [2]. The presence of the straight *U*-type steps is crucial to this anomalous behavior. It is clear that $T_g = 300$ °C provides enough thermal energy for deposited Si atoms to migrate to the nearest step edges since there is no nucleation in the middle of the terrace. It also appears in Fig. 3 that each cluster extends to the lower step edge. Thus, a likely explanation is that the nucleation of a very small amount of Si atoms on the transition region of the lower terrace can precede the growth of the clusters at the upper terrace. Ehrlich-Schwoebel barrier may contribute further to the confinement of clusters at the upper edge, but further study is necessary to draw a definite conclusion [19].

In summary, we have studied the very initial stage of Si growth on a well-defined, straight step-edge Si(111) surface. The single adatom rows and 1D cluster arrays are self-assembled along the step edges at $T_g = 400$ °C and 300 °C, respectively. The wire formation is allowed due to the presence of the transition region at the lower edge of the $U(2,0)$ step, while the 1D cluster array is formed by a different mechanism. We expect that isotopic versions of these nanostructures—$^{29}$Si nanowires and/or 1D nanoclusters on the vicinal $^{28}$Si(111) surface by Si isotope MBE [21]—will serve as basic building blocks for future development of silicon nuclear-spin quantum computers [8–10].

We would like to acknowledge H. Tochihara, S. Hasegawa, and I. Matsuda for fruitful discussions, and J. Lue and E.E. Haller for helpful discussions and reviewing. This work is supported in part by a Grant-in-Aid for Scientific Research in Priority Area "Semiconductor Nanospintronics" No.14076215.

FIG. 1 (color online). STM images before and after homoepitaxial growth. (a) Straight step edges before growth. Sample bias voltage $V_s$ is +1.0 V. (b) Si nanowires self-assembled along the lower step edges with $T_g$ = 400 °C and $w$ = 2$b$. $V_s$ = +1.5 V. (c) 1D arrays of Si nanoclusters self-assembled along the upper step edges with $T_g$ = 300 °C and $w$ = 7$b$. $V_s$ = +1.5 V.

FIG. 2 (color). (a) Magnified STM image of the straight step edge, taken with $V_s$ = +1.5 V. (b) A corresponding ball-and-stick model. (c), (e) Magnified STM images of the single adatom row, taken with $V_s$ = +1.5 V and $V_s$ = –1.5 V, respectively. (d) A ball-and-stick model corresponding to (c). The area of (e) corresponds to the rectangle in (c). Green and red arrows indicate the positions of the step edge before and after the wire formation, respectively, and blue arrows point at the dimer wall closest to the step edge on the lower terrace. The region between the green and blue arrows is referred to as the "transition region". In the models, green, blue, and red circles indicate atoms in the upper terrace, the lower terrace, and the self-assembled wire, respectively. The top-layer atoms on each region are indicated by filled circles, while underlying atoms are indicated by open circles with decreasing diameters for deeper ones. To highlight the rest atoms in the grown wire, their circles are exceptionally filled with pink.

FIG. 3 (color online). Magnified STM image of the 1D array of Si nanoclusters, to show that they are preferentially nucleated on the unfaulted halves at the upper step edge. $V_s$ = +1.5 V.

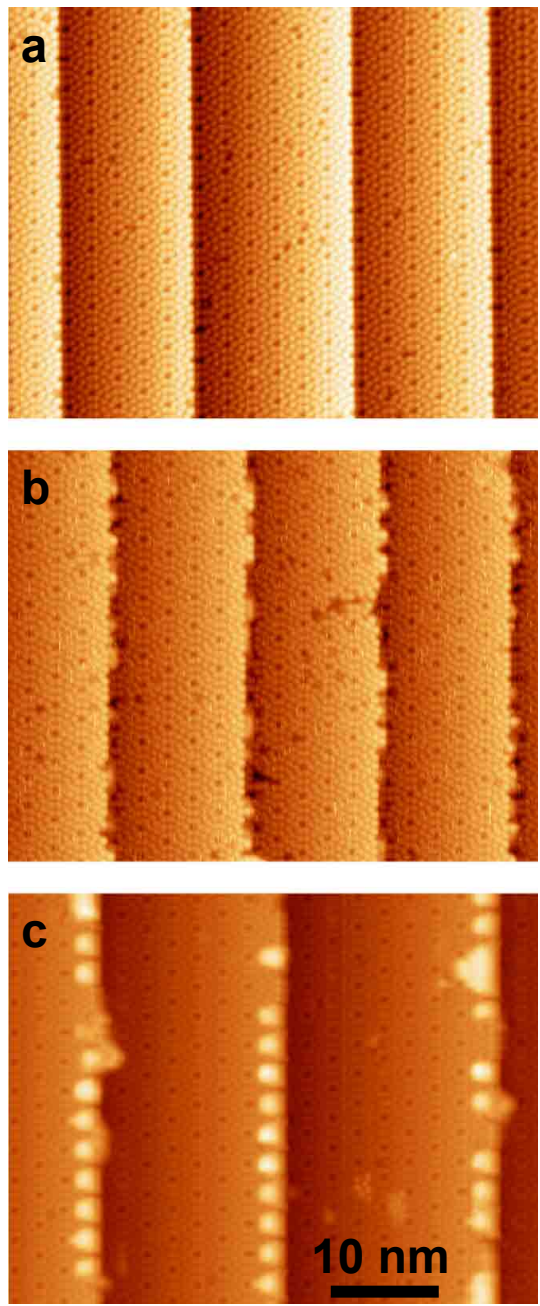

FIG. 1: T. Sekiguchi *et al*. Submitted to Phys. Rev. Lett.

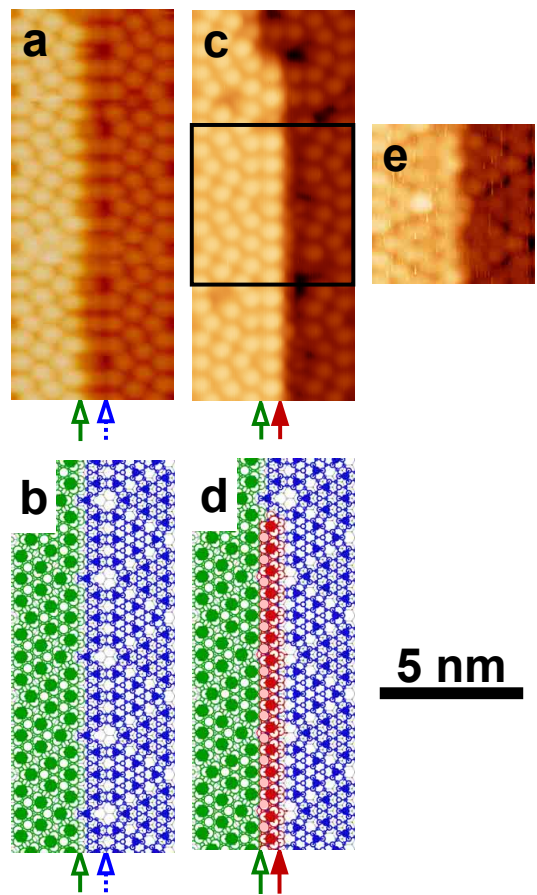

FIG. 2: T. Sekiguchi *et al*. Submitted to Phys. Rev. Lett.

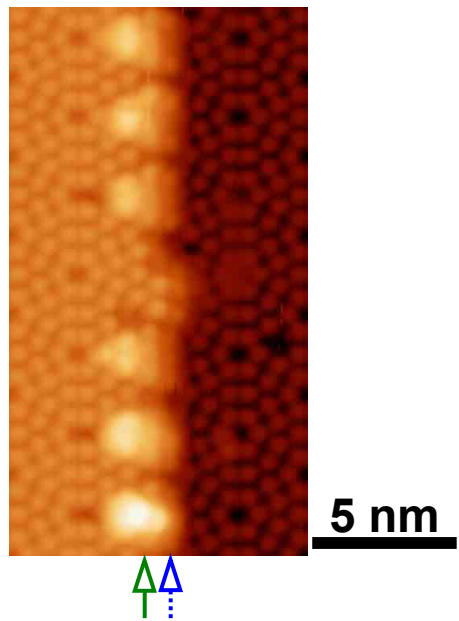

FIG. 3: T. Sekiguchi *et al*. Submitted to Phys. Rev. Lett.